\documentclass[aps,prl,twocolumn,superscriptaddress,floatfix,citeautoscript]{revtex4}
\usepackage{times}
\usepackage{graphicx}
\usepackage{dcolumn}
\usepackage{bm}
\usepackage{color}
\usepackage{amssymb}   
\newcommand {\beq} {\begin{equation}}
\newcommand {\eeq} {\end{equation}}
\newcommand {\bqa} {\begin{eqnarray}}
\newcommand {\eqa} {\end{eqnarray}}
\newcommand{\nbr} {\ensuremath{\langle ij \rangle}}

\begin{document}

\title{From immunity to sudden death: Effects of strong disorder in strongly correlated superconductors}

\author{Debmalya Chakraborty}
\affiliation{Indian Institute of Science Education and Research-Kolkata, Mohanpur, India-741246}

\author{Rajdeep Sensarma}
\affiliation{Department of Theoretical Physics, Tata Institute of Fundamental Research, Mumbai, India-400005}

\author{Amit Ghosal}
\affiliation{Indian Institute of Science Education and Research-Kolkata, Mohanpur, India-741246}

\begin{abstract}

We investigate the effect of strong disorder on a system with strong electronic repulsion. In absence of disorder, the system has a d-wave superconducting ground-state with strong non-BCS features due to its proximity to a Mott insulator.
We find that, while strong correlations make superconductivity in this system immune to weak disorder, superconductivity is destroyed efficiently when disorder strength is comparable to the effective bandwidth.
The suppression of charge motion in regions of strong potential fluctuation leads to formation of Mott insulating patches, which anchor a larger non-superconducting region around them. The system thus breaks into islands of Mott insulating and superconducting regions, with Anderson insulating regions occurring along the boundary of these regions. Thus, electronic correlation and disorder, when both are strong, aid each other in destroying superconductivity, in contrast to their competition at weak disorder.
Our results shed light on why Zinc impurities are efficient in destroying superconductivity in cuprates, even though it is robust to weaker impurities.

\end{abstract}

\maketitle

Strong inter-particle interactions and strong inhomogeneous potentials both tend to localize fermions.
 Strong repulsion can result in complete suppression of charge motion at commensurate filling, leading to a Mott insulator \cite{RevModPhys.70.1039}, while strong disorder, causes decoherence of fermions triggering formation of Anderson insulators~\cite{Anderson58}. There is some evidence that weak disorder in presence of strong interactions~\cite{0034-4885-67-1-R01,Punnoose289,PhysRevLett.81.4212,PhysRevLett.78.3943,PhysRevB.30.527} as well as strong disorder in presence of weak interactions \cite{Basko20061126} compete with each other, but the question of strong disorder in presence of strong repulsion remains unresolved. This is not merely an issue of theoretical interest, since the complex interplay of electronic interactions and disorder in two-dimensional (2D) materials is often crucial to understanding novel phenomena~\cite{Hirschfeld09, Balatsky06, Patrik14, Balents10,PhysRevLett.115.077001,Mirandabook12} beyond the standard paradigm of Fermi liquid and BCS superconductivity.

A prototype of strongly interacting electronic systems is the cuprate high $T_c$ superconductors (HTSC), which are antiferromagnetic Mott insulators at half-filling (one particle per site) and show d-wave superconductivity for a range of doping. In this paper, we will consider the effect of strong disorder on the strongly interacting d-wave superconducting (SC) state proximal to the Mott insulator. Our key findings are: (i) While the presence of strong correlations makes superconductivity robust to weak disorder, at large disorder comparable to bandwidth, superconductivity is rapidly suppressed. (ii) At large disorder, Mott insulating patches anchor a surrounding region akin to Anderson insulator. With increasing disorder strength, these islands grow at the expense of local superconductivity. Thus at large disorder, strong correlation and strong potential fluctuations help each other in bringing about the sudden death of superconductivity. The three distinct regions leave clear signatures in the local density of states. Our results shed light on why small concentration of strong substitutional impurities in cuprate superconductors (e.g. Zinc substituting copper in YBCO) degrades $T_c$ drastically, while superconductivity remains robust to weaker impurities~\cite{Garg08}.

The study of disorder in d-wave SC phase has a long history~\cite{Abrikosov61, Maki01, Hirschfeld00}, with early treatment within a Hartree-Fock-Bogoliubov inhomogeneous mean field theory (IMT)~\cite{Ghosal00, Hirschfeld00}, which ignores the effects of strong electronic correlations. Strong Mott correlations and consequent projection of the low energy Hilbert space into states with no double occupancies~\cite{ANDERSON1196,PhysRevB.38.931,Paramekanti04,RevModPhys.78.17} are however crucial to understanding the non-BCS character of the d-wave SC state in cuprates. A semi-analytic approach, where effects of projection are kept in terms of renormalization of Hamiltonian parameters, is the Gutzwiller approximation~\cite{Zhang88}, which is known~\cite{Rajdeep05} to match the more sophisticated Monte Carlo results~\cite{Paramekanti04} for the homogeneous system. This approach is easily extended to inhomogeneous situations to get a renormalized inhomogeneous mean field theory (RIMT) ~\cite{Garg08,Fukushima09,DC14,Vlad16}, which tries to capture effects of both strong correlations and disorder in the system.

A surprising result of RIMT~\cite{Garg08,Fukushima09,DC14,Vlad16} is that in-spite of the d-wave nature of the order parameter, strong correlations make superconductivity robust up to moderate disorders. This is ascribed to the electronic repulsions that modify the hopping amplitudes based on local density and smear out charge accumulation near deep potential wells, leading to a much weaker effective disorder. The natural question arises: How does Anderson localization~\cite{Anderson58} set in? Further, does presence  of strong repulsion, the largest energy scale in the problem, compete with or aid the localization of the electronic wave-function for large disorder strengths? 

In RIMT, strong interactions are treated non-perturbatively to obtain a low energy effective Hamiltonian and disorder potential is added to this description afterwards,  which fails to account for the fact that if the potential difference across a bond is much larger than the hopping scale, it is energetically unfavourable for the electron to hop across that bond. In this paper, we consider an extension of RIMT which builds in the absence of hopping across bonds with large potential difference across them, and thus includes the Anderson mechanism of localization in a more direct way. This approximation, called c-RIMT, allows us to smoothly interpolate between a robust SC at weak disorder to a patchy system of Mott and Anderson-like insulator at larger disorder strengths and shows the transition from immunity to sudden death of SC in the system.

\begin{figure}[t]
\includegraphics[width=0.45\textwidth]{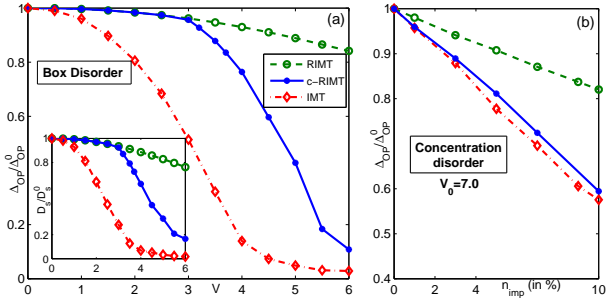}
\caption{(a) $\Delta_{\rm OP}$ (normalized by its value at $V=0$): Solid blue line represents c-RIMT results which crosses over from its robust nature (for $V \leq 3$) to its crashing down for $3 \lesssim V \lesssim 6$. $\Delta_{\rm OP}$ continues to be far less sensitive to $V$ within RIMT (dashed green line), and shows a continuous fall starting right from small $V$ within IMT scheme (dot-dashed red line). Thus, c-RIMT results interpolate between RIMT and IMT findings. Inset: Superfluid stiffness, $D_s$, shows a similar trend of $\Delta_{\rm OP}$, leading to its rapid destruction beyond $V_c$, whereas it depletes only weakly for small $V$. (b) Evolution of $\Delta_{\rm OP}$ with concentration of strong impurities in c-RIMT method shows impressive match with IMT findings, rather than with RIMT results.
}
\label{fig:odlro}
\end{figure}

{\it Model and methods}----
We work with the disordered Hubbard model on a square lattice
\beq
{\cal H}=-t\sum_{i,\sigma, \atop \delta=\hat{x},\hat{y}} (c_{i \sigma}^{\dagger}c_{i+\delta \sigma}+{\rm h.c.}) + U\sum_i n_{i \uparrow}n_{i \downarrow}\sum_{i\sigma} (V_i-\mu)n_{i\sigma}
\eeq
Here, $t$ and $U$ denote hopping and onsite repulsion energies respectively, $c^\dagger_{i\sigma}$ and $n_{i\sigma}$ are the creation and number operators for electrons on site $i$ with spin $\sigma$.  We work with the strong coupling parameter $U =12t$, and choose an average filling of $\rho=0.8$, so that the homogeneous system is a d-wave superconductor even in the presence of strong correlations. The non-magnetic impurity potential $V_i$ is taken from a uniform distribution between $-V/2$ and $V/2$. We emphasize that while we focus on $V \gtrsim t$, we always consider $V\ll U$, so that the projection constraints remain valid in our system~\cite{Rajdeep16}.

At low energies, the homogeneous Hubbard model can be reduced to an effective $t-J$ model in the subspace where double occupancies are projected out through an appropriate Schrieffer Wolff transformation~\cite{PhysRevB.37.9753} about a local Hamiltonian. A similar procedure can be carried out in the disordered model by including the disorder potential non-perturbatively in the local Hamiltonian. In this case the potential difference across a bond provides an additional energy scale (other than $U$), which determines the effective Hamiltonian on that link. At weak potential difference across a link \nbr, $\Delta V_{ij}=|V_i-V_j| <V_c$, this gives the standard $t-J$ model with a super-exchange scale $J_{ij} = (4t^2/U)(1-\Delta V_{ij}^2/U^2)^{-1}$~\cite{RevModPhys.78.17} on that link. However, for $\Delta V_{ij} > V_c \sim t$, hopping on the corresponding link is frozen as the electrons pay a large potential energy cost to hop across this bond. The bond is effectively {\it cut} as far as transport is concerned, although the electrons interact with the corresponding superexchange scale, as mentioned above. The critical disorder $V_c \approx 2.8t$, is determined by balancing the kinetic energy gain with the potential energy loss for a single-impurity problem with a local potential $V$.
We solve our modified ``$t-J$" Hamiltonian within RIMT formalism, where $t_{ij} \rightarrow g^t_{ij} t_{ij}$ and $J_{ij} \rightarrow g^s_{ij}J_{ij}$ with
$g^t_{ij} = 2[x_ix_j/(1+x_i)(1+x_j)]^{1/2}$ and $g^s_{ij}=4/(1+x_i)(1+x_j)$.
Here $x_i$ is the local hole doping which is determined self-consistently together with a Fock shift ($\tau_{ij}$) and a d-wave pairing amplitude ($\Delta_{ij}$) on each bond (See supplementary material (SM) for details). In this paper, we will present results on $30 \times 30$ lattice (with a repeated zone scheme \cite{PhysRevB.66.214502,DC14} used on $12 \times 12$ unit cells for better resolution and statistics, see SM). We will express all energies in units of $t$.

{\it Demise of superconducting correlations.}---- To look at the robustness of SC, we study the off diagonal long range order, $\Delta_{\rm OP}^2=\lim_{|i-j| \to \infty} F_{\delta, \delta'}(i-j)$, where $F_{\delta, \delta'}(i-j)=\langle B_{i \delta}^{\dagger} B_{j \delta'} \rangle$. Here, $B_{i \delta}^{\dagger}=( c^{\dagger}_{i \uparrow} c^{\dagger}_{i + \delta \downarrow} + c^{\dagger}_{i + \delta \uparrow} c^{\dagger}_{i  \downarrow})$ is the singlet Cooper-pair creation operator on the bond $(i,i+\delta)$.
Conventional Abrikosov-Gorkov theory~\cite{Abrikosov61} for a d-wave superconductor demands $\Delta_{\rm OP}$ to rapidly degrade with $V$. Incorporating fluctuations in pairing amplitude, an IMT calculation slows down such decay~\cite{Ghosal00}, nevertheless, its demise still occurs as shown in Fig.~\ref{fig:odlro}(a) (dot-dashed line).
Including strong correlation effects within RIMT, where $\Delta_{\rm OP} \sim \sum_{\nbr} g^t_{ij} \Delta_{ij}$, is known~\cite{Garg08,DC14} to make superconductivity rather immune to disorder, as plotted in Fig.~\ref{fig:odlro}(a) (dashed line). In this case, large local densities approaching unity, lead to a decrease in the kinetic energy around those sites due to the renormalization factors. This non-linear effect creates a repulsive potential and leads to a weak effective disorder in these systems, thereby making $\Delta_{\rm OP}$ robust. 

Our c-RIMT calculation is identical to RIMT for $V \le V_c$ as there are no cut-bonds. However, in the range $V_c \le V \le 6.0$ upto 60\% of kinetic links are frozen, and $\Delta_{\rm OP}$ depletes by nearly 90\% . In this case, local potential wells, where the density reaches nearly unity, are also accompanied by large potential differences in bonds connected to the wells, i.e. to frozen bonds. Thus the renormalization of the disorder potential around these wells are absent, leading to formation of Mott insulating sites which anchor regions of large differences in site energies on neighboring sites -- reminiscent of Anderson insulator, causing rapid destruction of superconductivity. We have checked that a change in $V_c$ merely gives a parallel shift to the trace $\Delta_{\rm OP}(V)$, without any qualitative modification.

The sudden demise of the superconducting correlations for $V>V_c$ is also signalled by the superfluid stiffness, $D_s$ (shown in the inset of Fig.~\ref{fig:odlro}(a)), which is defined by,
\beq
\frac{D_s}{\pi}=\langle -k_x \rangle - \Lambda_{xx}(q_x=0,q_y \rightarrow 0,\omega=0),
\label{eq:supd}
\eeq
where, $k_x$ is the kinetic energy along the $x$-direction and $\Lambda_{xx}$ is the long wavelength limit of transverse (static) current-current correlation function~\cite{PhysRevB.47.7995}. We find that the behavior of $D_s$ shows strong parallel with the $V$-dependence of $\Delta_{\rm OP}$, confirming the trends seen in the nature of $\Delta_{\rm OP}$.

We also examined a different model of disorder: Randomly located impurities of strength $V_0$ on $n_{\rm imp}$ fraction of sites, which is more relevant to Zn doping of cuprate high-temperature superconductors (HTSC)~\cite{Pan99,Yazdani99,Hirschfeld15}.
Such impurities suppress SC dramatically~\cite{Nachumi96}. Zinc impurities in cuprate HTSC have traditionally been treated as strong repulsive potential \cite{PhysRevB.67.094508,Ziegler00,PhysRevB.84.184511}, although recent work has shown these impurities to be attractive \cite{Balatsky06,Hirschfeld15}. We show here that for large repulsive $V_0=7.0$ ($>V_c$), the $n_{\rm imp}$-dependence of $\Delta_{OP}$ follows the weak coupling IMT behavior, rather than the strong coupling RIMT trend~\footnote{In both IMT and RIMT, $\rho_{\rm imp} \rightarrow 0$ for large $V_0=7.0$ and the links connecting to impurities have no particle-particle or particle-hole amplitude due to the large difference of disorder on them.}, as shown in Fig.~\ref{fig:odlro}(b). In contrast, a healthy $\Delta_{\rm OP}$ persists up to a considerably large $n_{\rm imp}$ (similar to RIMT) for weaker $V_0 \lesssim 3$~\cite{DC14,Garg08,Vlad16}. In the SM, we show that for strongly attractive $V_0=-4.0$, the behavior interpolates between the RIMT and IMT findings. Our results thus explain the loss of superconductivity in HTSC with Zinc impurities for both the repulsive and the attractive impurity strength.

\begin{figure}[t]
\includegraphics[width=0.45\textwidth]{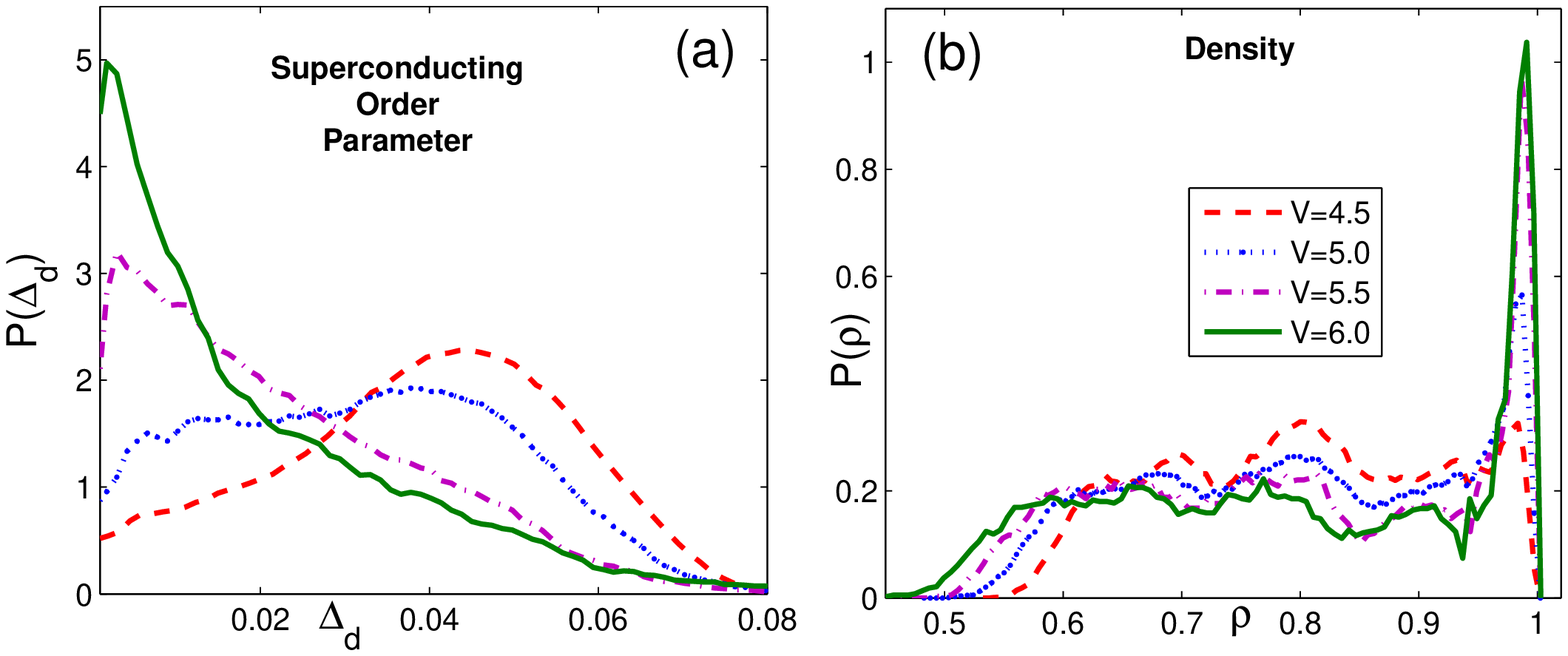}
\includegraphics[width=0.45\textwidth]{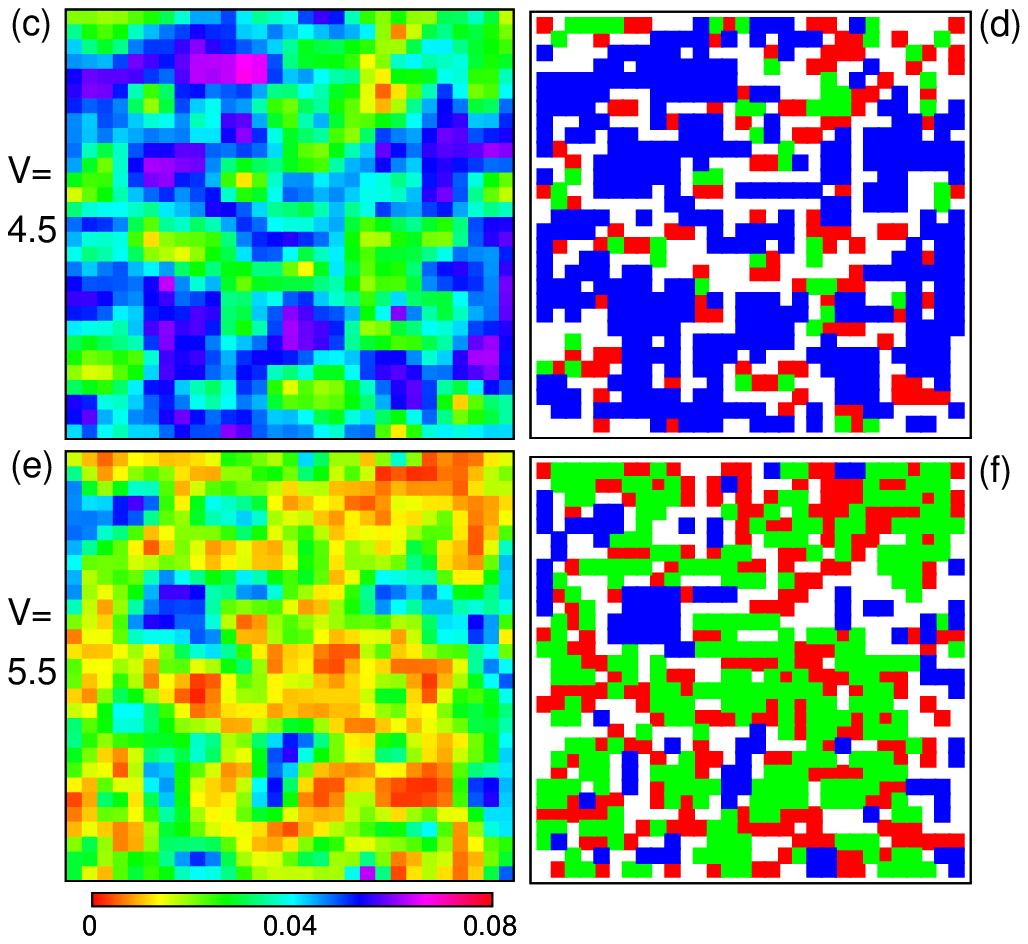}
\caption{Distribution of (a) $\Delta_{d}$ and (b) $\rho$ for $V=4.5-6.0$. $P(\Delta_d)$ broadens with increasing $V$ with a peak at $\Delta_{d}=0$, while Mott-sites show up as a peak in $P(\rho \rightarrow 1)$ that sharpens with $V$. Color density plots of $\Delta_{d}$ on lattice are shown for (c) $V=4.5$, and (e) $V=5.5$. Increasing $V$ shrinks the blue regions forming superconducting `islands' in the matrix of non-superconducting regions. The plots in (d) and (f) shows the cross-correlation of density and order-parameter. In this map, (i) blue dots correspond to superconducting regions ($\Delta_d > 0.7 \Delta_d(V=0)$), (ii) red dots correspond to Mott regions ($\rho>0.98$), and (iii) green dots correspond to non-SC, non-Mott region ( $\Delta_d < 0.3 \Delta_d(V=0)$ and $\rho < 0.95$). Note that green regions are always nucleated around red regions.
}
\label{fig:op}
\end{figure}

{\it Distribution of local order parameters}----
The picture we have painted above, that at large $V$, disorder and interaction aid each other in killing SC, is validated when we look at the distribution of local order parameter, $\Delta_d(i)=\frac{1}{4} \sum_{j=n.n.} (-1)^{\delta_{j,i \pm \hat{y}}} g^{t}_{ij} \Delta_{ij}$ and the local density $\rho_i$ as a function of $V$. This is plotted in Fig.~\ref{fig:op}(a) and (b) for several values of $V>V_{c}$. $P(\Delta_d)$ broadens with increasing $V$ developing a peak at $\Delta_d(i) \approx 0$, similar to IMT results ~\cite{Ghosal98,Ghosal01}, and in stark contrast to RIMT results~\cite{DC14}, where the distribution form narrow bands. However, unlike IMT, the importance of correlations become evident from Fig.~(\ref{fig:op}(b)  where the density distribution starts growing a strong peak at $\rho \approx 1$, indicating the importance of formation of locally Mott insulating regions in the demise of superconductivity. 

Our c-RIMT calculations, however afford us a granular view of the system in terms of spatial arrangements of different types of regions. To see this, the spatial distribution of $\Delta_d(i)$ are shown in Fig.~\ref{fig:op} (c) and (e) for $V=4.5$ and $V=5.5$, which shows the formation of superconducting and non-superconducting islands, with non-SC islands growing with disorder. However, a clearer picture emerges if we cross-correlate the spatial distributions of order parameter and local densities. The easiest way to present this data is to divide the sites into three representative classes (i) Mott insulating sites, where local density $\rho(i) > 0.98$ (ii) Superconducting sites where $\Delta_d(i) > 0.04$ and (iii) sites with low order parameter ($\Delta_d(i) < 0.02$) and density not close to $1$ ($\rho(i) < 0.95$), the non-SC, non-Mott sites, which we will interpret as consisting of Anderson insulating patches. Fig.~\ref{fig:op} (d) and (f) present this cross-correlated data corresponding to the order parameter maps in Fig.~\ref{fig:op} (c) and (e) for $V=4.5$ and $V=5.5$ respectively. Here the superconducting sites are colored blue, the Mott insulating sites are colored red, while the non-SC as well as non-Mott insulating sites are colored  green. Fig~\ref{fig:op} (d) clearly shows that Mott insulating sites act as anchors around which the insulating patches nucleate. With increasing disorder, these ``Anderson insulating'' patches (green) form a network connecting the Mott sites. The fraction of both the red and green sites grow with disorder. Thus, `island-formation' in a d-wave SC, where both the electronic repulsion and disorder are strong, is more subtle than in a s-wave superconductor \cite{Ghosal98}. The Mott correlations and disorder potential aid each other in the limit of strong disorder to localize the electrons and kill superconductivity.

\begin{figure}[t]
\includegraphics[width=0.5\textwidth]{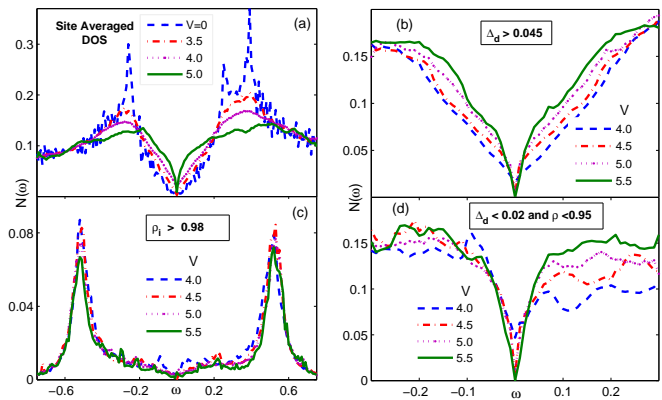}
\caption{(a) Site-averaged DOS, $N(\omega)$ showing filling up of mid-gap states for strong $V$ and depletion of coherence peaks. (b) (c) (d): $N(\omega)$ in (b) superconducting regions, (c) Mott-clusters and (d) Anderson insulating region.
The superconducting region shows depleted coherence peak at $\omega =\pm 0.26$ for all $V$ and Mott-clusters show a spin gap at $\omega \sim J_{\rm eff}/2 \approx \pm 0.58$. The Anderson insulating region shows an otherwise flat DOS, except for a sharp gap feature at $|\omega| \leq 0.05$.
}
\label{fig:dos}
\end{figure}

{\it Local Density of states.}----
The three types of patches discussed above leave their signatures in the local density of states (DOS) at these points, $N(i,\omega)= g^t_{ii} \sum_n \left\{ |u_{i,n}|^2\delta(\omega-E_n) + |v_{i,n}|^2\delta(\omega+E_n) \right\}$ \cite{DC14,Garg08}, where ($u_{i,n},v_{i,n}$) are the local Bogoliubov wavefunctions corresponding to energy eigenstates with energy $E_n$. In Fig.~\ref{fig:dos} (a), we plot the DOS averaged over all sites in the system. At weak disorder, the V-shaped low energy DOS is robust to disorder, which mainly affects the coherence peak at the  gap edge. At larger disorder, superconducting coherence peaks deplete significantly  and there is a filling of the d-wave gap, although a narrow gap exists even at strong disorder strength of $V=5.0$. 

The local density of states, averaged over the sites belonging to the three categories mentioned above show distinct features of their own. In the superconducting regions (Fig.~\ref{fig:dos} (b)), we find that the density of states continue to show the low energy V shaped feature characteristic of d-wave superconductors. As disorder is increased, the slope of the DOS with energy steepens at very low energy, indicating transfer of spectral weight from high energies at the gap edge to the low energies. This is consistent with the fact that at strong disorder, the quasiparticles become heavier with $V$ and hence disorder averaged effective velocities $v_F$ and $v_\Delta$ decreases. So the low energy spectral weight $N(\omega) \sim \omega/v_Fv_\Delta$ grows with $V$. In the Mott regions (Fig.~\ref{fig:dos} (c)), there is a clear gap in the low energy DOS with particle-hole symmetric sharp peaks at $\omega=\pm0.58$, the location and lineshape of which is robust to changes in $V$. This is because the Mott-clusters are described by an effective Heisenberg Hamiltonian for localized spins in a basis without any double occupancy. The difference between the singlet and the triplet energies in this model is $\Delta_{\rm spin} \sim J_{\rm eff}$, the exchange coupling of that Hamiltonian and is independent of the disorder. This is the scale that shows up in the DOS of the Mott regions, further confirming our association of these sites with Mott insulating patches. While disorder indeed generates these patches, $\Delta_{\rm spin}$ must be independent of $V$ once such a model is in place, which is consistent with our findings. In the third region (Fig.~\ref{fig:dos} (d)), we find a DOS which is flat at the energy-scale of superconducting coherence peaks (similar to Anderson insulators, and hence the name), but features a tiny gap at very small $|\omega| \lesssim 0.05$. In these regions, the low energy DOS first shows signs of gap filling at intermediate $V$, but as the disorder increases, there is a depletion of spectral weight at low energies, leading to a fully formed gap by $V=5.0$. 
A thin gap in disordered dSC had already been discussed in weak-coupling theories~\cite{Nersesyan94, Senthil99, Hirschfeld00, Ziegler00, Ghosal00}. In addition, Coulomb repulsions are known to open up a gap in disordered systems~\cite{AA79,AA85,Efros75,Prabudhya08}.
Our results emphasize the role of strong correlations in the low-energy spectrum, the details of which will be addressed elsewhere.

{\it Conclusion.}----
We have studied the effects of strong potential disorder on strongly interacting d-wave superconducting states in proximity to a Mott insulator. Using a c-RIMT method, which explicitly freezes hopping on bonds with large potential difference, we find that, while strong correlations effectively compete against weak disorder to make superconductivity immune to disorder, at large disorder strengths, correlations and disorder aid each other leading to sudden demise of superconductivity. This is facilitated by formation of Mott insulating patches, which anchor Anderson insulating patches around them. Quantum phase fluctuations in real materials, which are beyond the scope of this study, are likely to bring quantitative changes in our descriptions, however, the evolution of intertwined regions will still survive and their distinct signatures in the local density of states, can be picked up by scanning tunnelling microscopy.

{\it Acknowledgements.}----
Authors acknowledge computational facilities at Dept. of Theoretical Physics, TIFR Mumbai and at IISER Kolkata. AG acknowledges P.~J. Hirschfeld for valuable discussions. DC acknowledges fellowship from CSIR (India).

\bibliographystyle{apsrev}
\bibliography{Draft.bib}

\pagebreak
\widetext
\clearpage 
~\vspace{2cm} 
\setcounter{equation}{0}
\setcounter{figure}{0}
\setcounter{table}{0}
\setcounter{page}{1}
\makeatletter
\renewcommand{\theequation}{S\arabic{equation}}
\renewcommand{\thefigure}{S\arabic{figure}}

\begin{center}
{\Large\bf Supplementary Material to ``From immunity to sudden death: Effects of strong disorder in strongly correlated superconductors"}
\end{center}

\vspace{2.5cm}

\section{Renormalized inhomogeneous mean field theory (RIMT)}\label{simpcri}

The Hubbard model, a minimal model to describe correlated systems, is given by, 
\begin{equation}
{\cal H}_{\rm Hubb}= -t \sum_{\langle ij \rangle \sigma} (c^{\dagger}_{i \sigma} c_{j \sigma}+h.c.)+U\sum_j n_{j \uparrow}n_{j \downarrow}.
\label{eq:Hubb}
\end{equation}
Here, $t$ is the hopping energy between nearest neighbor sites $i,j$ indicated as $\langle ij \rangle$ and $U$ is the onsite interaction mimicking screened Coulumb electronic repulsion.
In the strongly correlated limit $U \gg t$, the standard Schrieffer-Wolff transformation on ${\cal H}_{\rm Hubb}$ yields an effective $t-J$ model in the low energy subspace:
\begin{eqnarray}
{\cal H}_{\rm t-J} & = & \sum_{\langle ij \rangle \sigma}-t (\tilde{c}^{\dagger}_{i \sigma} \tilde{c}_{j \sigma}+h.c.)+\sum_{\langle ij \rangle}J \Big(\tilde{\mathbf{S}}_i.\tilde{\mathbf{S}}_j-\frac{\tilde{n}_i\tilde{n}_j}{4}\Big) \nonumber \\
&-&\frac{J}{4}\sum_{\langle ijm \rangle, \sigma \atop m\neq i} (\tilde{c}^{\dagger}_{i \sigma} \tilde{n}_{j \bar{\sigma}} \tilde{c}_{m \sigma}-\tilde{c}^{\dagger}_{i \sigma} \tilde{c}^{\dagger}_{j \bar{\sigma}} \tilde{c}_{j \sigma} \tilde{c}_{m \bar{\sigma}}+h.c.) 
\label{eq:tJfull}
\end{eqnarray}
where all the terms up to ${\cal O}(t^2/U)$ are kept, $J=4t^2/U$ and $\tilde{c}_{i\sigma}=c_{i \sigma}(1-n_{i\bar{\sigma}})$ is the annihilation operators in the `projected space' prohibiting double-occupancy at the site $i$. The term in the second line is the three-site term involving three nearest neighbors $\langle ijm \rangle$. Though this term contributes energy of the order of $J$, it is already verified in Ref.~\cite{DC14} that three-site terms do not introduce any new qualitative physics -- even in the presence of disorder. So, for the sake of simplicity, we do not consider them for all our calculations here. We introduce disorder by redefining ${\cal H}_{\rm t-J}$ to ${\cal H}_{\rm t-J}+ \sum_{i\sigma} (V_i-\mu)n_{i\sigma}$, where $V_i$ is the (non-magnetic) impurity potential at site $i$ and $\mu$ is the chemical potential that fixes the desired average density of electrons. Using Gutzwiller approximation, $t-J$ model in Eq.~(\ref{eq:tJfull}), ignoring three-site terms, and in the presence of disorder, can be written as,
\begin{equation}
{\cal H}_{\rm t-J}  =  -t \sum_{\langle ij \rangle \sigma} g^{t}_{ij} ({c}^{\dagger}_{i \sigma} {c}_{j \sigma}+h.c.)+\sum_{\langle ij \rangle}J \Big(g^{s}_{ij} {\mathbf{S}}_i.{\mathbf{S}}_j-\frac{{n}_i{n}_j}{4}\Big) + \sum_{i\sigma} (V_i-\mu)n_{i\sigma}
\label{eq:tJgut}
\end{equation}
where $g^{t}_{ij}$ and $g^{s}_{ij}$ are the Gutzwiller renormalization factors defined in the main text.
We solve ${\cal H}_{\rm t-J}$ using inhomogeneous mean field theory with the local parameters, which need to be calculated self consistently, being $\rho_i\equiv \langle c_{i \downarrow}^{\dagger} c_{i \downarrow} \rangle_0 + \langle c_{i \uparrow}^{\dagger} c_{i \uparrow} \rangle_0$, $\Delta_{ij} \equiv \langle c_{j \downarrow} c_{i \uparrow} \rangle_0 + \langle c_{i \downarrow} c_{j \uparrow} \rangle_0$, and $\tau_{ij} \equiv \langle c_{i \downarrow}^{\dagger} c_{j \downarrow} \rangle_0 \equiv \langle c_{i \uparrow}^{\dagger} c_{j \uparrow} \rangle_0$. Here $\langle \dot \rangle_0$ implies expectation with respect to the ground state wave function  in the Hilbert space with no double occupancy constraint. We will use $U=12$ for RIMT calculations. For a justified comparison, we choose $U=3.3$ in the IMT calculations ($g^{t}_{ij}=1$ and $g^{s}_{ij}=1$), which yields the same uniform d-wave gap from RIMT at $V=0$.

\section{Modifications in the Schrieffer-Wolff transformation: \lowercase{c}-RIMT}

In the limit $V \lesssim t$, when the disorder is weak compared to the other energy scales of the problem (such as $U$ and $t$), the inclusion of disorder by promoting ${\cal H} \rightarrow {\cal H}_{\rm t-J}+ V_{\rm dis}$ (standard implementation in RIMT) is valid. However, if $t\ll V \ll U$ (the regime of our interest), Schrieffer-Wolff transformation must be carried out directly on the {\it disordered} Hubbard model. In our regime of interest ($U \gg t$ and $U \gg |V_i-V_j|$ for all $\langle ij \rangle$), performing such Schrieffer-Wolff transformation on, ${\cal H}^{\rm dis}_{\rm Hubb}=-t\sum_{\langle ij \rangle \sigma} (c_{i \sigma}^{\dagger}c_{j \sigma}+{\rm h.c.}) + U\sum_j n_{j \uparrow}n_{j \downarrow}+ \sum_{i\sigma} (V_i-\mu)n_{i\sigma}$ we obtain,
\begin{eqnarray}
&&{\cal H}_{\rm eff} = \sum_{\langle ij\rangle \sigma} \Big[ \Theta \left(|V_{i}-V_{j}|-V_c \right) {\cal H}_{\rm A}(i,j) + \Theta \left( V_c - |V_{i}-V_{j}|\right) {\cal H}_{\rm B}(i,j) \Big] \nonumber \\
&+&\sum_{\langle ij\rangle} J_{ij} \Big(\tilde{\mathbf{S}}_i.\tilde{\mathbf{S}}_j-\frac{\tilde{n}_i\tilde{n}_j}{4}\Big) +\sum_{i}(V_{i}-\mu)n_{i} 
\label{eq:effective}
\end{eqnarray}
where ${\cal H}_{\rm A}(i,j)=-t (\tilde{c}^{\dagger}_{i \sigma} \tilde{c}_{j \sigma}+h.c.)$ and ${\cal H}_{\rm B}(i,j) = t^2 (V_i-V_j)^{-1} (n_{j \sigma}-n_{i \sigma})$. Here, $J_{ij}=4t^{2}U [U^{2}-(V_{i}-V_{j})^{2}]^{-1}$, which, in the limit of weak disorder, expectedly leads to $J_{ij}=J=4t^2/U$.
Thus the effective low energy sector is decided individually for each link of the lattice. Hoppings that do not even change the number of double occupancy can be prohibited on certain bonds in the lattice depending on the disorder difference ($|V_i-V_j|$) between the sites $\langle ij \rangle$ connecting them. Bonds with $|V_i-V_j|< V_c$ are governed by ${\cal H}_{\rm A}(i,j)$ and bonds with $|V_i-V_j|> V_c$ are governed by ${\cal H}_{\rm B}(i,j)$, where $V_c$ is the critical disorder strength decided energetically. This Hamiltonian implies that the bonds with $|V_i-V_j|> V_c$ will be governed by no direct hopping but will have higher order exchange processes. In the actual calculation, the heaviside $\Theta$-functions in Eq.~(\ref{eq:effective}) are replaced by smoother $n_{F}(x/\Gamma)$ where $n_{F}(\frac{x}{\Gamma})=[1+exp(x/\Gamma)]^{-1}$ to account for a realistic smoothness in kinetic `bond-cutting'. It is ensured that these two functions have similar qualitative outcome. 

Achieving self consistency in the presence of strong correlations and for strong disorder (both RIMT and c-RIMT) is difficult. As the strength of the disorder is increased, the density of sites with highly attractive disorder potential tend to reach $\rho \approx 1$ and the self consistency on this sites becomes progressively difficult due to resulting divergences in the Hartree-shift of the chemical potential involving derivative of $g^{t}_{ij}$. To tackle this problem, we have added an additional term $U\sum_j n_{j \uparrow}n_{j \downarrow}$ and treat them at the Hartree level. We have used the same value of Hubbard $U$ for this purpose. This is implemented for all the calculations (i.e., IMT, RIMT and c-RIMT) in the main text for justified comparison between them.

\section{Determination of $V_{c}$}

\begin{figure}[h]
\includegraphics[width=0.6\textwidth]{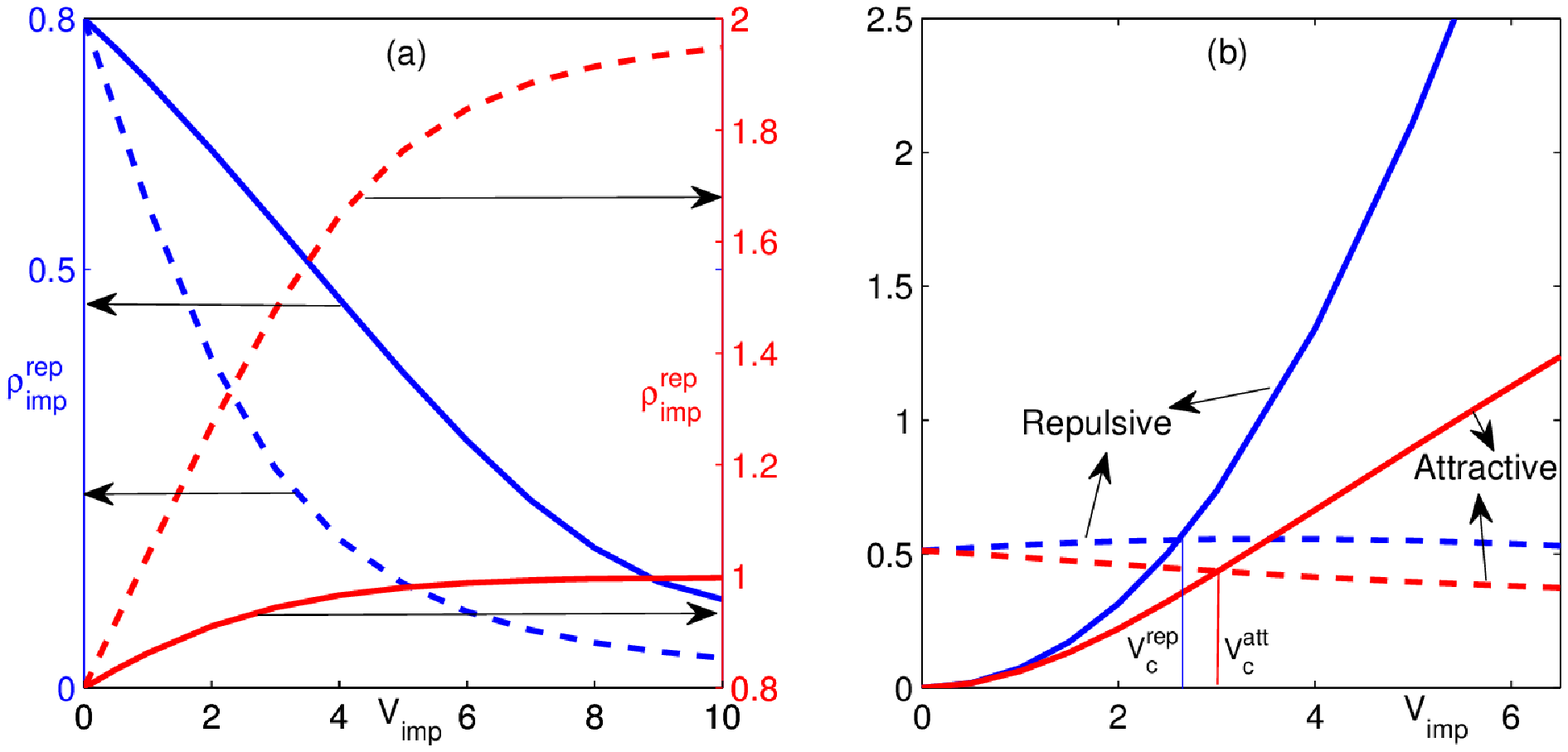}
\caption{(a) $\rho_{\rm imp}$ for different $V_{\rm imp}$ of single impurity. Left axis gives the densities on repulsive impurities ($\rho_{\rm imp}^{\rm rep}$) with dashed line being the results in IMT and solid line in RIMT. In the both the cases, $\rho_{\rm imp}^{\rm rep}$ approaches zero as $V_{\rm imp}$ is increased. Right axis shows the corresponding densities ($\rho_{\rm imp}^{\rm att}$) for attractive single impurity. $\rho_{\rm imp}^{\rm att}$ approaches the value 2 for IMT (dashed line), whereas the value is restricted to 1 in RIMT (solid line) by construction of the theory. (b) Comparison of different energy scales in RIMT to determine the criterion of `cutting' the bonds. Solid lines are the disorder energies which increase with increasing impurity strength. Dashed lines are the absolute values of kinetic energy gains which remain more or less constant in the scale of disorder energy for high $V_{\rm imp}$. The crossing point of these two curves decides $V_c$.
}
\label{fig:criterion}
\end{figure}
Let us consider the single impurity problem and imagine increasing the strength ($V_{\rm imp}$) of a sole impurity (at $r_{\rm imp}$) in an otherwise homogeneous background. In IMT, increasing the strength of repulsive and attractive impurity eventually leads to the density of electrons on the impure site, $\rho_{\rm imp}=0$, or $2$, respectively (dashed lines in Fig.~(\ref{fig:criterion}a)). In RIMT, strong repulsive and attractive impurity sites become insulating due to Anderson physics ($\rho_{\rm imp} \approx 0$) and Mott physics ($\rho_{\rm imp} \approx 1$), respectively. Thus, for large $V_{\rm imp}$ (and {\it irrespective of its sign}), the impurity site must be kinetically decoupled from its neighbors. 

We illustrate this by showing a crossover of energy gain for kinetic delocalization and disorder energy cost (defined as $V_{\rm imp}|\rho_{\rm imp}- \rho |$) for a single impurity within RIMT formalism. The constant part ($\rho V_{\rm imp} $) can be approximately thought of (at least for weak $V_{\rm imp}$) as the change in $\mu$ due to the impurity. For our parameters, the crossover takes place at $V_{\rm c}=2.6$ for repulsive and $V_{\rm c}=3.0$ for attractive $V_{\rm imp}$ (Fig.~(\ref{fig:criterion}b)).
For an attractive
impurity, the density fluctuation is strongly suppressed as it takes the impurity site close to the
Mott limit (with $\rho=0.8$). The value of disorder energy cost, subsequently, is much smaller for attractive impurity leading to $V_{\rm c}^{\rm att}>V_{\rm c}^{\rm rep}$.
RIMT method, that renormalizes local hopping and thereby homogenizing electron density, does not incorporate this important physics of hopping prohibition based on impurity strengths. 
Based on this comparison of a single impurity problem within RIMT, we choose $V_c=2.8$ -- the average value for attractive and repulsive impurities. We also choose $\Gamma=0.03$ for our c-RIMT calculations.

\section{$\Delta_{\rm OP}$ with attractive concentration disorder}

\begin{figure}[t]
\includegraphics[width=0.48\textwidth]{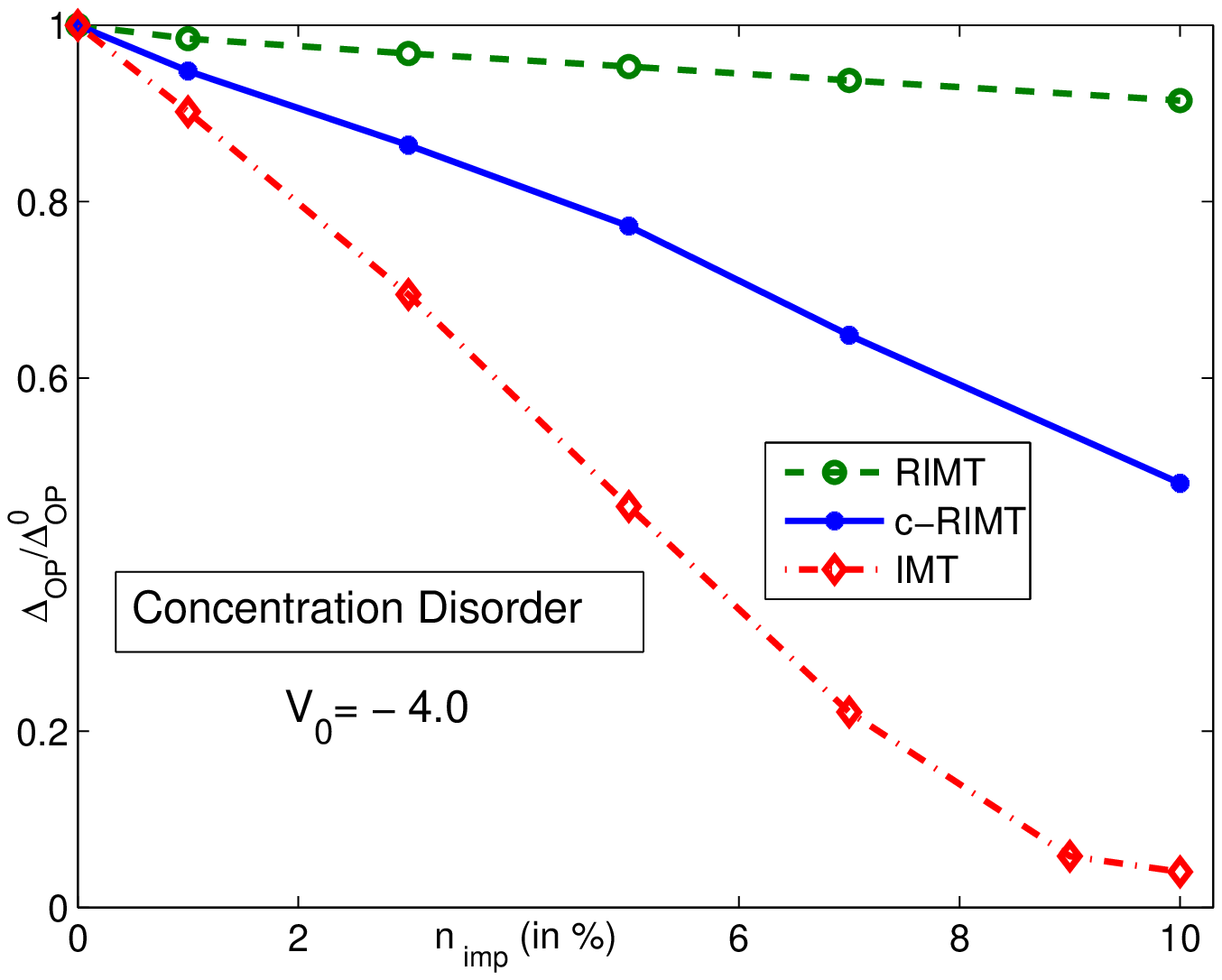}
\caption{$\Delta_{\rm OP}$ with attractive concentration impurities ($V_0=-4.0$) too degrade $\Delta_{\rm OP}$ in c-RIMT (solid line) by $60\%$ for $n_{\rm imp}=10\%$, which is a weaker fall than IMT results (dot dashed line), and RIMT (dashed line) maintains strong superconducting correlations.
}
\label{fig:concattodlro}
\end{figure}

The results of attractive concentration disorder ($V_0 < 0$) is also interesting and shown in Fig.~(\ref{fig:concattodlro}). While qualitative findings are similar to the repulsive case (Fig.~\ref{fig:odlro}(b) in main text), quantitative physics are different. This emphasizes that for $\langle \rho \rangle =0.8$, the role of attractive and repulsive impurities differ in the presence of Mott correlations -- a feature prevalent in the box-disorder.
With attractive impurities ($V_0=-4.0$), $\Delta_{\rm OP}$ in c-RIMT suffers significant fall by $n_{\rm imp}=10\%$, but the RIMT shows its robustness -- a result quite similar to those in Fig.~\ref{fig:odlro}(b). The IMT results, on the other hand, falls very rapidly to zero by $n_{\rm imp}=9\%$. This is because $\rho^{\rm IMT}_{\rm imp} \rightarrow 2$ and as $n_{\rm imp}$ increases, the sites without impurities turn locally overdoped to maintain desired density. As a result, $\Delta^{IMT}_{\rm OP}$ falls rapidly in IMT in contrast to c-RIMT that only allows $\rho^{\rm c-RIMT}_{\rm imp} \lesssim 1$. As we increase the strength of attractive impurities ($|V_0|$), $\rho^{\rm RIMT}_{\rm imp} \rightarrow 1$. This causes difficulty in numerical convergence for $V_{0}<-4.0$ in RIMT calculations due to associated diverging contributions in the Hartree-shift. Note that even when $\rho^{\rm c-RIMT}_{\rm imp} \rightarrow 1$, numerical stability prevails due to kinetic freezing of links connecting to the impurities removing any divergences in the Hartree-shift.

\section{Calculating Superfluid stiffness}

The defining characteristic of a superconductor lies in the Meissner \cite{book:17888} effect, which is quantified by the stiffness of the ground state wave function to an externally applied phase twist. This rigidity translates into its finite superfluid stiffness, $D_s$, which is proportional to the superfluid density. Within the framework of linear response theory, the Kubo formalism derives superfluid stiffness as the following,
\begin{equation}
\frac{D_s}{\pi}=\langle -k_x \rangle - \Lambda_{xx}(q_x=0,q_y \rightarrow 0,\omega=0),
\label{eq:supd}
\end{equation}
where $k_x$ is the kinetic energy along the $x$-direction (which is the diamagnetic contribution to $D_s$) and $\Lambda_{xx}$ is the long wavelength limit of transverse (static) current-current correlation function \cite{PhysRevB.47.7995}. $\Lambda_{xx}$ is calculated by Fourier transforming the impurity averaged Matsubara Green's function;
\begin{equation}
\Lambda_{xx}(\mathbf{q},i\omega_n)=\frac{1}{N}\int_0^{1/T}d\tau e^{i\omega_n\tau}\langle j_x^p(\mathbf{q},\tau)j_x^p(-\mathbf{q},0)\rangle,
\label{eq:lambda}
\end{equation}
where $j_x^p(\mathbf{q})$ is the paramagnetic current and $\omega_n=2\pi nT$ ($n$ is a positive integer). The corresponding Gutzwiller factors for $\langle -k_x\rangle$ and $\Lambda_{xx}(\mathbf{r_i},\mathbf{r_j},\tau)$ are $g^{t}_{i,i+x}$ and $g^{t}_{i,i+x}g^{t}_{j,j+x}$ respectively. Pure BCS superconductors offer no paramagnetic response from the current-current correlation, leading to perfect diamagnetism. Disorder generates such response, turning it into a non-superconductor when this paramagnetic response equals the diamagnetic one.

\section{Repeated Zone Scheme}

We extended our numerical calculations to larger system (called a supercell) containing identical copies of smaller unit cells each of size $30 \times 30$. Translational operator, which repeats the unit cells to construct the supercell with periodic boundary condition, commutes with the Hamiltonian. So, the eigenstates of this translational operator can be used to block diagonalise the Hamiltonian of the supercell, following similar ideas behind Bloch's theorem. Such a method is commonly known as `repeated zone scheme' (RZS) \cite{PhysRevB.66.214502}. Here, we have used a supercell containing up to $12 \times 12$ unit cells. RZS calculations are numerically inexpensive compared to the BdG calculations on corresponding larger system. Since the disorder profile of the unit cell is repeated in the supercell, we need to average over large number of disorder configurations for statistical inferences minimizing the impurity-impurity correlations. We have averaged over up to 15 disorder realisations.

We used RZS in the calculation of DOS by considering a supercell containing $12 \times 12$ unit cells. Effective size of the supercell is now $360 \times 360$ generating more number of states with in the band width, which produces a denser spectrum in DOS. While RZS improves the resolution in DOS, we verified that the distributions of all order parameters as well as their spatial structures remain unchanged by going from one unit cell to $12 \times 12$ unit cells. Besides, for the calculation of superfluid stiffness, obtaining the $q_y \rightarrow 0$ limit of $\Lambda_{xx}$ is limited by the number of $q_y$ values available on a $30 \times 30$ system. It is thus essential to obtain data on larger systems using RZS for an appropriate $q_y \rightarrow 0$ extrapolation \cite{DC14}. A significant numerical demand still limits the number of unit cells up to $3 \times 3$ in this case.

\end{document}